# Sensitive Pictures

## Emotional Interpretation in the Museum

Steve Benford

Mixed Reality Lab, University of Nottingham, Nottingham, United Kingdom, steve.benford@nottingham.ac.uk

Anders Sundnes Løvlie

Digital Design Department, IT University of Copenhagen, Copenhagen, Denmark, asun@itu.dk

Karin Ryding

Digital Design Department, IT University, Copenhagen, Denmark, kary@itu.dk

Paulina Rajkowska

Dept of Informatics and Media, Uppsala University, Uppsala, Sweden, paulina.rajkowska@im.uu.se

Edgar Bodiaj

Mixed Reality Lab, School of Computer Science, University of Nottingham, Nottingham, United Kingdom, edgar.bodiaj@nottingham.ac.uk

Dimitrios Paris Darzentas

Mixed Reality Lab, University of Nottingham, Nottingham, Nottinghamshire, United Kingdom, Dimitrios.Darzentas@nottingham.ac.uk

Harriet R Cameron

Mixed Reality Lab, University of Nottingham, Nottingham, Nottinghamshire, United Kingdom, Harriet.Cameron@nottingham.ac.uk

Jocelyn Spence

Mixed Reality Lab, University of Nottingham, Nottingham, Nottinghamshire, United Kingdom, Jocelyn.Spence@nottingham.ac.uk

Joy Egede

Mixed Reality Lab, University of Nottingham, Nottingham, Nottinghamshire, United Kingdom, Joy.Egede@nottingham.ac.uk

Bogdan Spanjevic

NextGame, Belgrade, Serbia and Montenegro, bogdan@nextgame.rs

Museums are interested in designing emotional visitor experiences to complement traditional interpretations. HCI is interested in the relationship between Affective Computing and Affective Interaction. We describe Sensitive Pictures, an emotional visitor experience co-created with the Munch art museum. Visitors choose emotions, locate associated paintings in the museum, experience an emotional story while viewing them, and self-report their response. A subsequent interview with a portrayal of the artist employs computer vision to estimate emotional responses from facial expressions. Visitors are given a souvenir postcard visualizing their emotional data. A study of 132 members of the public (39 interviewed) illuminates key themes: designing emotional provocations; capturing emotional responses; engaging visitors with their data; a tendency for them to align their views with the system's interpretation; and integrating these elements into emotional trajectories. We consider how Affective Computing can hold up a mirror to our emotions during Affective Interaction.

# 1 Introduction

The nature of emotion and its role in our interaction with computers has become a topic of great interest to HCI. Emotion has come to be recognised as an integral part of "third wave" user experience [47]. Affective Computing, in which artificial intelligence strives to recognise human emotions, promises experiences that respond to our emotions [57]. Affective Interaction employs digital technology to engage humans with their own emotions [11]. The relationship between these overlapping paradigms is unclear and a topic of ongoing debate within the community [36].

In turn, contemporary museums are seeking to engage emotion in new approaches to interpreting their exhibits, branching out from traditional information giving towards more personal, subjective and emotional visitor experiences [3,41,66,69]. Many are turning to digital technologies to achieve this, employing mobile experiences and interactive installations to overlay layers of interpretation on their exhibits as part of an increasing "hybridization" of museums [10,22,51]. Museums therefore offer a powerful context for engaging the public in emotional interpretation and through this, contributing to the wider debate about digital technologies and emotional user experiences.

Our paper reports the design of a visiting experience that aimed to engage visitors in an emotional interpretation of artworks. This involved designing digital technologies to stimulate emotional responses before then enabling reflection on these. The experience added layers of overtly provocative emotional content to exhibits; captured both self-reported and machine-interpreted (using computer vision) measures of emotional response; and presented these back to visitors as data souvenirs.

We report a design-led and research-in-the-wild engagement in which, over a period of eighteen months, we co-created this emotional visiting experience with the Munch Museum, home to a major collection of Edvard Munch's famously emotionally charged paintings. We describe how 132 members of the public experienced our design, drawing on interviews with 39 of them to uncover their experience and attitudes. This reveals how our design tended to provoke strong emotional responses; how it could be difficult for both human and machine to capture this; but how even limited data presented on a souvenir postcard stimulated visitors to interpret their emotions post-hoc, often with a marked tendency to agree with the system's analysis. This leads us to present further designs for richer and more nuanced visualizations of emotional response that might be more widely deployed around the museum. We also reflect on Affective Computing's potential role as a mirror, albeit a somewhat distorted one, to human emotions.

## 2 RELATED WORK

The question of "what are emotions?" is both deeply fascinating and profoundly challenging, having engaged thinkers from many disciplines, from Neuroscience mapping of emotion to complex interactions among the brain's various systems, to Psychology's various taxonomies of emotions, to perspectives from the Arts and Humanities that see emotions as being socially constructed, interpreted by humans and artistically provoked. These have long been the subject of extensive academic debate [e.g. 16], a debate that has been further fuelled by AI's attempts to "measure" emotions by analyzing various physiological signals [50]. We cannot do full justice to this debate in the space available, but aim to give a brief overview of research that has shaped our approach to designing emotional visiting experiences.

### 2.1 Emotions in computing, design and HCI

Over the last two decades, affect and emotion has become a topic of great interest within various fields. This trend started sometime in the late 1990s, influenced by a wave of research on emotions and affect within Psychology, Neurology, Medicine, and Sociology [16,40,43]. However, this "affective turn" has been far from homogeneous [81], with researchers adopting distinctly different theoretical underpinnings and methodological approaches. This is equally true when it comes to research on affect and emotion related to technological systems and their design in HCI.

Perhaps the best-known direction related to HCI is "Affective Computing" after a ground-breaking book by Rosalind Picard [57]. This grounds emotions in biological processes, aiming to establish digital technologies that capture, analyze and adapt to various physiological signals of emotion [57:1]. Affective Computing typically employs digital technologies to measure emotions and affective states, including sensors that capture facial expressions, body postures, gestures, heart rate, EEG and other potential signals. Analysis of these is often grounded in taxonomies of emotion from Psychology such as the Circumplex Model of valence and arousal [60], Ekman's taxonomy of universal emotions [23] and models of how facial expressions convey inner emotions [24]. Affective computing research has explored health applications as part of "Behaviomedics" [78], including assessing mental health conditions [58], pain [21], post-traumatic stress disorders [49] and autism [59]. Other applications include measuring engagement-related social cues, e.g. frustration, delight and confusion, during learning [17].

The challenges of trying to map emotions in this way, as if they come in discrete packets, templates or programmes, each with its own embodied signature or brain/body routine, have been articulated by a great deal of social research [e.g. 44,80,81]. In response, an alternative approach called Affective Interaction has emerged that draws upon phenomenology to view emotion as something constructed in the interactions between people and between people and machines [11,12,27,34,38,75]. As Boehner et al. underline, "instead of sensing and transmitting emotion, systems should support human users in understanding, interpreting, and experiencing emotion in its full complexity and ambiguity" [12:275 emphasis added]. Affective Interaction typically draws on qualitative methods such as interviews, self-reporting of emotional states [39], and cued-recall – a form of situated recall to elicit information about users' affective states during the use of a system [8]. It may include the creation of affective loops in which systems both respond to and generate affect as part of embodied interaction [35]. Affective Diary by Lindström et al. [45] encouraged participants to interpret ambiguous representations of their own sensor data which has been captured and uploaded via their mobile phone. Leahu, Schwenk and Sengers [42] give participants access to recordings of their objective signals (in this case, measurement of arousal patterns through skin conductivity) to allow them to take control of the emotional interpretation.

Höök contrasts Affective Computing and Affective Interaction with a third perspective, Technology as Experience, a holistic perspective in which emotions are viewed as core to user experience and also "inseparable from the intellectual and bodily experiences" [36]. This builds on John Dewey's seminal work on aesthetic experience [18] and subsequent research in HCI that has explored the design of aesthetic and emotional interactive experiences, including that of McCarthy and Wright [47] and Gaver, who argues that emotion cannot be teased out as a separate facet of experience design [27].

## 2.2 Emotional experiences in museums

Museums have a unique ability to give visitors emotional, restorative and transformative experiences. This is made possible because museums are liminal spaces set apart from ordinary life [5]. From visitor research it is known that the contemplative, spiritual and restorative qualities of museum visits affect a great number of visitors, and can even be the driving motivation for some visitors dubbed "Rechargers" [25]. However, even though museums and heritage sites are clearly places where people go to feel, explore and to "manage" their emotions in a safe environment [4,68], emotions have often been treated with suspicion in museum and heritage studies, and even as somehow "dangerous" when it comes to achieving a balanced understanding of the importance of the past in the present [70]. Debate on the value of emotional experiences in museums has, for example, flared up in response to an increasing emphasis (often seen as being populist and puerile by traditionalists) on tactile, visual, and aural aids to interpretation which are often digitally mediated [70,71]. In turn, researchers within museum and heritage studies have engaged with the role of affect and emotion in shaping people's relations to cultural heritage. This includes the identification of "deep empathy" as key to triggering a response that engages the imagination in such a way that visitors start to question what they know and understand [67,68,82].

HCI also has a longstanding engagement with museums. In contrast to early technological interventions that were most often preoccupied with digital information delivery to visitors [32,56,74,79], in more recent years, following the overall academic trend, we have seen an increase of interest in visitor's affective engagement during museum visits. Examples include the work by Fosh et al. [26] and Hazzard et al. [33], which both use mobile technology to deliver interactive music and narratives to emotionally enhance a visit to a sculpture garden. Spence et al. [72] introduce the practice of gifting as a way of mediating intimacy and creating affective engagements as part of a museum visit. This can be compared to the work by Ryding et al. [61,63] in which visitors are given the tools to create intimate and affective experiences for each other as they explore the museum together.

Another relevant thread of research concerns providing visitors with tangible "data souvenirs" of their visits. Aipperspach et al. [1] propose that tangibility is a useful tactic for designing reflective technologies and creating physical booklets as tangible souvenirs of domestic experiences. Petrelli et al. [55] brought this approach into museums, showing that personalized tangible souvenirs generated from data captured during a visit can engage visitors with online content afterwards. In other work, they describe how automatic text generation techniques can create personalized postcards [53]. Looking beyond HCI, one of the most notable examples of creating emotional data souvenirs lies in the work of the artist Christain Nold and his practice of biomapping and emotional cartography [52]. Nold established a participatory methodology in which people, equipped with a combination of GPS and GR ("sweat") sensors, explored towns and cities, capturing data that was subsequently reflected back to them on cartographic emotion maps so as to elicit their responses to the urban environment. Also relevant, though not directly focused on museum visits, is a study in which the riders of thrilling amusement rides were kitted out with personal telemetry systems that streamed video, heart rate and skin resistance data (potential signals of emotional arousal) to a watching audience. The study noted a marked tendency for riders to "perform" their emotional journeys throughout the ride, narrating their feelings of fear and excitement, and noted that the act of being publicly "kitted out" may have stimulated this storytelling behavior as much as the data itself [64].

In conclusion, while affect and emotion are high on the research agenda of museums and heritage studies, these topics are contested, in particular in relation to electronically mediated interpretations of cultural heritage. Researchers in HCI have mostly responded to this challenge of engaging visitors emotionally through the use of technology by focusing on studies relying on qualitative data of their museum experiences.

## 3 APPROACH AND DESIGN PROCESS

We adopted the methodology of performance-led research in the wild [6] in which artists and researchers co-create cultural experiences that are then studied "in the wild" of live public deployments. This is a design- and practice-led approach that involves reflection on both the design process and subsequent experience of a public audience in order to generalize design knowledge. Our experience, Sensitive Pictures, was co-created by Bogdan Spanjevic, an artist/designer (and author of this paper) who has a background in theater and also leads the professional user experience design company NextGame; the Munch Museum at Oslo where it was deployed and studied as a public experience; and our research team. The Munch Museum is an art museum located in Oslo, Norway, that is dedicated to the works and life of the Norwegian artist Edvard Munch who was renowned for his emotionally resonant paintings. His most famous painting, The Scream, is prominently displayed in the museum. Sensitive Pictures was part of a larger project to develop new interactive museum experiences that engaged multiple museums across Europe.

We broadly follow an Affective Interaction approach, engaging visitors in interpreting their own emotional responses, but we incorporate elements of both the Affective Computing and Technology as Experience approaches. Our specific design-orientation was one of "subjective-objectivity" that, according to Leahu, Schwenk and Sengers [42], highlights the role of human interpretation in negotiating the relationship between the objective signals and subjective experiences of emotions. We directly drew on Affective Computing technologies in the form of computer vision, specifically facial expression recognition systems, to automatically generate emotion data, embedding these into the wider visitor experience. Our approach also reflects elements of Technology as Experience in that it interleaves emotional provocation, data capture and reflection into an extended visiting experience, though it differs by clearly singling out emotion as a separate facet of experience to be reflected upon.

Our design process began in May 2018 and ran through Summer 2019. A first workshop explored available emotion detection technologies, leading to the decision to work with the BlueMax computer vision technology, a state of the art Affective Computing research tool at the time [50]. We concluded that it should be able to reliably detect three relevant affective states: enjoyment, engagement and frustration. Creative workshops employed techniques from improvisational theater to envisage new kinds of emotional visitor experience; six of Munch's paintings were selected based on their emotional resonance, and participants staged drama exercises in front of these to envisage related emotional scenarios. The full visitor experience, including scripted stories, emotion capture techniques and visualizations, was then iteratively developed, including two rounds of testing on site, with the final version that we describe below being publicly deployed at the Munch Museum between 28–31 August 2019.

132 public visitors completed the full experience, while a further 65 engaged with part of the experience. We captured system logs of their interactions with the mobile web app (described in section 4.2), including self-reported emotion data, alongside measures of signals of emotion from our computer vision system. We conducted semi-structured interviews with 39 participants, who were approached after finishing the experience. Interviews were recorded and transcribed and the results were analyzed. Initial discussions among the research team identified preliminary themes which were then used to systematically code the transcribed interviews before then being further discussed and refined. The self-reported emotion data was used to generate a series of post-hoc visualizations that we report in the discussion below.

Our study was granted ethical approval by the University of Nottingham. Part of our approach to the sensitivities of handling personal data was to explicitly give participants the choice of either donating their data to the museum or withholding it after they had completed the experience, in contrast to giving blanket consent to hand over their data before they fully understood its nature and their own response. Out of the 132 who completed the experience only 1 declined to donate their data, which was then deleted. For the further 65 who participated but did not fully complete the experience, all data was deleted as they did not reach the end station and so could not actively consent to donating their data after completion.

As a final note, following our deployment and study, the museum decided to press ahead with a more permanent visitor experience focused on emotional interpretations via audio narratives coupled with a personalized music experience. They are currently seeking funding to develop an emotional visitor experience that, in their words, will "spark emotions in the user, establish connections between user and art based on feelings, shared associations and memories and emotional responses."

# 4   THE DESIGN OF SENSITIVE PICTURES

The final design of Sensitive Pictures as presented to the public comprised four parts: (i) onboarding; (ii) using a mobile web app to listen to emotional stories while standing in front of selected paintings before then self-reporting one's emotional response; (iii) a video installation that delivers a simulated "phone conversation" with a fictionalized Munch during which computer vision technology detects social signals of emotional response from the visitor's facial expressions; and (iv) a souvenir postcard that summarizes the visitor's "emotion data" for them to reflect on.

## 4.1 Onboarding

Visitors entering the museum were invited to try out Sensitive Pictures by members of the research team. Participants were given a card that could be used to open the web app on their phones. They were informed what kinds of data the experience would capture and that they would be given the opportunity at the end to decide whether to donate their data to our research or to have it deleted. The card provided a unique three-digit visitor identifier code and contained an embedded RFID chip to uniquely identify each visitor at the later video installation. Once set up, the visitor was able to freely navigate the museum in their own time, following their own trajectory.

## 4.2 Mobile web app

The mobile web app presented visitors with a choice of six emotions, laid out in a grid: "love", "self-confidence", "passion", "fear", "sadness" and "obsession" (Figure 1). On choosing one of these, the screen displayed a corresponding painting and instructed the visitor to locate it in the museum. Once it was found, the visitor pressed "play" to hear an audio recording with three parts: a dramatic fictional story offering an emotional interpretation of the painting, some factual information about the painting, and a question about their own emotions. For example, on choosing "love", the visitor would be asked to find Munch's painting Vampire and would be invited to listen to a story which begins as a dialogue between a female and a male voice set against a musical soundscape designed to enhance its emotional tone:

> Woman: He doesn't love me.
> Man: I gave her all my love.
> Woman: He is never at home! I miss him.
> Man: She'll drain all the life from me! As if she is going to eat me.
> Woman: Look at my arms, hugging this man. It is love!
> Man: It's pain. Look at my neck!
> Woman: He can hide in my hair. I will kiss him, and all his troubles will be gone.
> Man: That's not hair. It's blood! And not a kiss – but a bite.

After the end of this dialogue, the music changed and a new voice addressed the listener in a factual tone, similar to an ordinary audio guide:

> Vampire was painted in 1893, and is considered today as one of Edvard Munch's most renowned motifs. The picture has not always been titled Vampire. Initially Munch gave it the name "Love and Pain", and it was actually one of Munch's good friends in Berlin, the Polish writer Stanislav Przybyszewski, who suggested Vampire as a better title.

Followed by this question:

> Now think about the most intense relationship you have been in. Describe how you feel about it: whether it is a pleasant or an unpleasant emotion, and how strong is that feeling.

We include recordings of all six audio stories in the supplementary materials so that the reader can better appreciate the nature of the visitor experience. All six narratives were co-created by the artist Bogdan Spanjevic and a curator at the museum, Nikita Mathias. The text about Vampire was inspired by Munch's own life, as explained by Spanjevic:

> Munch had a mostly tormentous love life and relationships that were both erotic, passionate, and painful, almost destructive. That ambivalence was maybe most present and visible in his passionate relationship with Tula Larsen that ended violently – by Tula shooting Munch with a pistol. Although he was just lightly wounded – he was very traumatized by that event. So, when he portrays love, he usually depicts complex multilayered relationships – i.e. torturer/victim, fear/passion, etc. "Vampire" is an obvious example of such an approach. [Spanjevic, personal communication]

After the audio finished, the screen displayed a text input field as well as three sliders, prompting the visitor to describe their emotions with words and by adjusting the sliders (Figure 2). These sliders represent the three dimensions of the "affective slider" model of Betella & Verschure [9] (which builds on Bradley and Lang's Self-Assessment Manikin model [14]): valence, arousal and control, chosen because it has been widely used in research on self-reporting emotions.

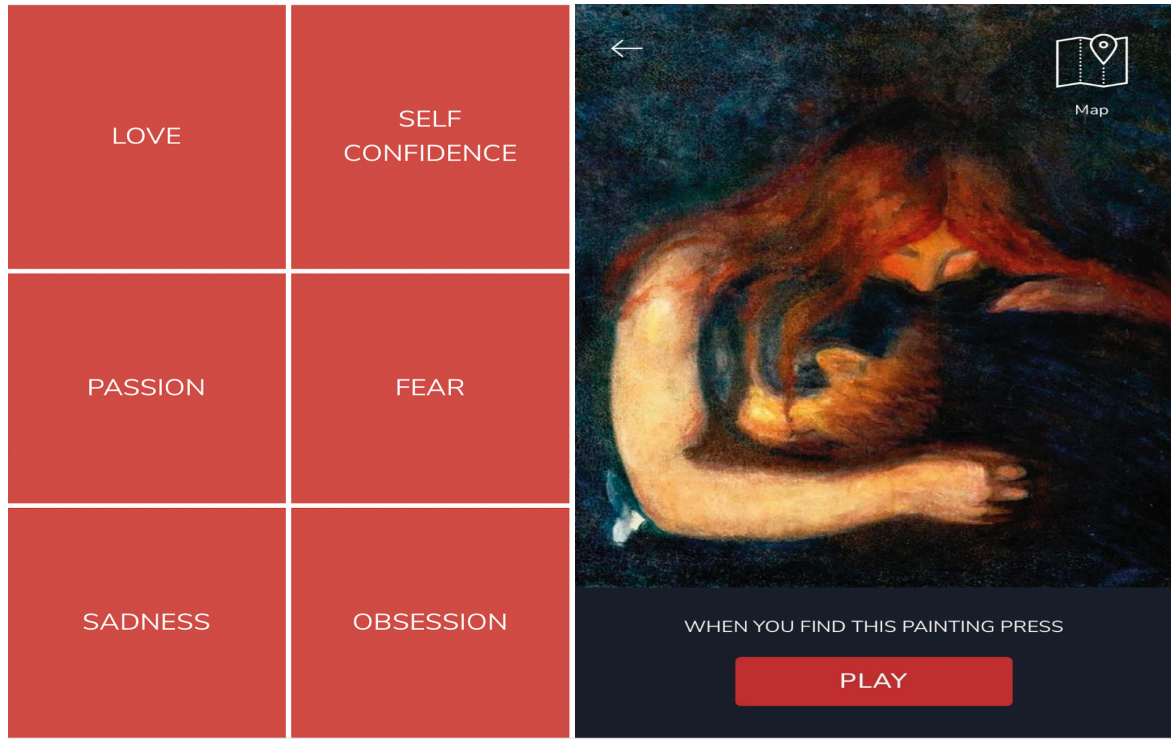

**Figure 1:** Instructions to choose an emotion and then find the corresponding painting.

**Figure 2:** Interface for self-reporting emotional responses.

## 4.3 Video installation

Early discussions considered how best to deploy the computer vision system in the museum. Using cameras on visitors' phones (assuming a mobile implementation of the software) would require them to point the camera at themselves throughout the experience, which might be distracting and socially awkward. Mounting cameras on the walls raised questions of practicality and privacy. We therefore opted to design a fixed installation in front of a monitor equipped with a camera that could be deployed away from the paintings in the main gallery space.

Once the visitor had finished using the web app, they were asked to return to the table by the entrance. There they were invited to step behind a divider screen and take a seat in front of a large monitor mounted on a painter's easel, next to a table with an old-fashioned telephone (Figure 3). They were instructed to insert their identifier card into a small box under the screen (where it was read by the RFID reader), at which point the telephone rang. Once the visitor answered the phone, the screen lit up and displayed one of Munch's self-portraits, while a voice on the phone greeted the visitor and presented itself as Edvard Munch.

The voice on the phone then talked to the visitor about their experience and emotions. Drawing on the data from the web app, "Munch" talked about the painting that the visitor had reported the strongest emotional reaction to as measured by their self-reported values of arousal. The self-reported measure of valence and control were used to choose in cases where two or more paintings shared the highest level of self-reported arousal. Munch would then address a positive or negative perspective relating to the chosen painting depending on the visitor's self-reported valence. Thus, the system selected from twelve pre-recorded videos (one emotionally positive and one emotionally negative for each painting). Munch would also ask the visitor questions about their experience, who was given the opportunity to respond aloud. We include one example of a video interview with Munch (Vampire, negative valence) as an accompanying video figure.

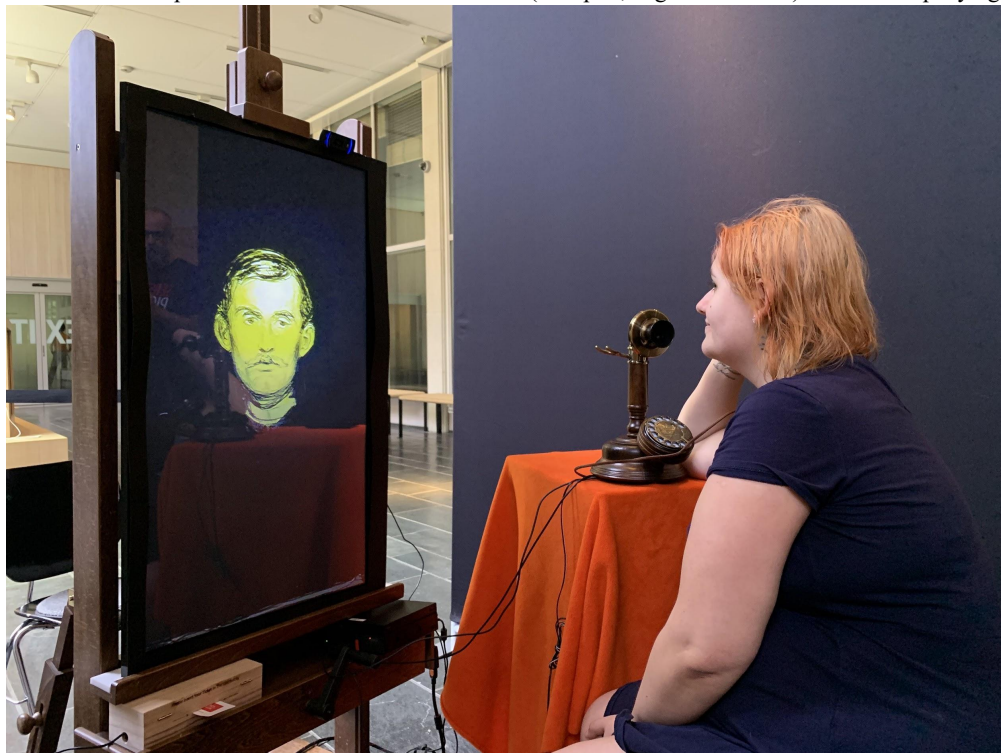

**Figure 3:** The phone conversation with Edvard Munch.

During this conversation, a camera mounted above the monitor captured video of the visitor's face which the BlueMax software would analyze in an attempt to classify their emotions. BlueMax employed image analysis to dynamically recognise the movements of key Facial Action Units (FAUs), based on Ekman and Frisen's [24] classification that has been widely adopted by the Computer Vision community. Early experimentation in situ determined that the system could reliably detect five key FAUs: cheek raiser (AU6), upper lip raiser (AU10), smile (AU12), dimpler (AU14), and chin raiser (AU17). To give greater accuracy, these were combined with estimations of head pose and movement [2]. BlueMax interpreted these outputs as signaling three affective states: degrees of enjoyment, engagement and frustration. This choice was constrained by what could be dynamically interpreted from the five reliable action units, though was also deemed to be relevant to visitors' possible emotional states in a museum. This interpretation was founded on prior cognitive-behavioral studies [17,30,46,48] that have identified the combinations of FAUs that make up common engagement-related behaviors. Our models here were defined as follows: Enjoyment is a combination of AU6 and AU12 intensities; Frustration combines AU10, AU17 and head activity; while engagement, which is typically viewed as a composition of other affective states in previous studies [17], was crafted as a combination of the five FAUs and head activity. In practice, the system often worked reliably when deployed in "the wild" of the museum, but did fail in some cases which meant that only portions of the experience with valid behavioral information were analyzed.

## 4.4 Souvenir postcard

Following the completion of the phone call with Munch, the visitor returned to the front desk. On handing their ID card over to the researchers, a souvenir postcard was printed. One side of the postcard showed the painting the visitor had reacted most strongly to. On the other side, there was a summary of their emotion data in three parts (e.g. Figure 4).

- TOP: A representation of their self-reported data from the valence and arousal sliders, mapped onto the Circumplex Model of emotion (valence on the x-axis, arousal on the y-axis). Each red dot corresponds to one of the paintings that they experienced, numbered in the order they visited them.
- MIDDLE: sentences summarizing their self-reported response to each painting, each in the form "At the [painting title] you felt X", where X is the participant's answer to the free-text question.
- BOTTOM: a sentence summarizing the output from the computer vision system, according to the three measures of Engagement, Enjoyment and Frustration. Each variable was mapped onto a four-point scale from "not at all" to "very" and then integrated in an explanatory sentence: "It seems like this experience has left you [not at all/a little/somewhat/very] frustrated, [not at all/a little/somewhat/very] engaged, and you enjoyed it [not at all/a little/somewhat/a lot]."

The final question on the postcard invites the visitor to reflect on whether this summary of their data does indeed represent how they felt.

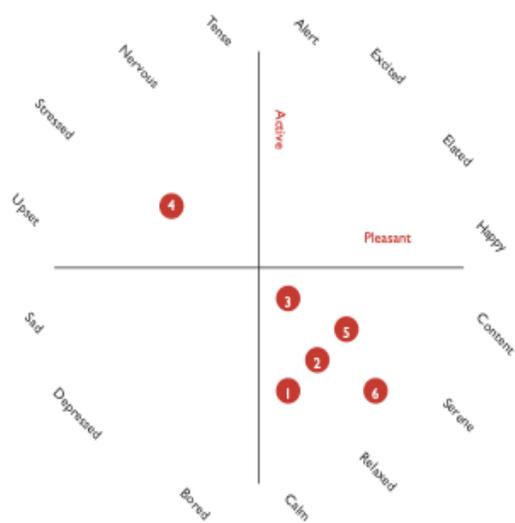

Figure 4: Postcard with visualization of emotion data from one user.

## 5 AUDIENCE EXPERIENCE OF SENSITIVE PICTURES

We now consider the public experience of Sensitive Pictures. On average, each of our 132 visitors spent 60 minutes completing the entire experience and experienced 5 paintings out of the 6 available. This resulted in a balanced choice of paintings/emotions, from the most popular (Vampire/love), which received 119 engagements, to the least (Christian Munch on the Couch/obsession), which received 98. Notably, Vampire was by far and away the most popular choice for the first painting to be experienced with 105 occurrences, compared to 15 for Self Portrait With Brushes, 6 for Madonna, 2 each for The Scream and The Sick Child and just 1 for Christian Munch on the Couch.
Considering which paintings provoked the strongest self-reported emotional responses (and so were chosen to be discussed in the phone conversation with Munch): The Sick Child got the strongest response from 37 participants, The Scream from 28, "Self-Portrait with Brushes" from 23, Vampire from 18, Christian Munch on the Couch from 17, and Madonna from 8. Figure 5 visualizes the spread of self-reported valence (x-axis) and arousal (y-axis) across the six paintings. For each painting, we display aggregate self-reports from one or more visitors as coloured circles plotted on the Circumplex Model. A circle appears if one or more visitors reported these levels of arousal and valence (they were reported as discrete levels, so visitors often reported the same values). The larger and redder the circle, the more visitors rated themselves with these particular values at this painting. These visualizations suggest that our paintings provoked varied emotional responses. We see, for example, that The Scream has a largely negative valence and somewhat excited emotional footprint, whereas "Self-Portrait with Brushes" has a predominantly positive and aroused footprint.

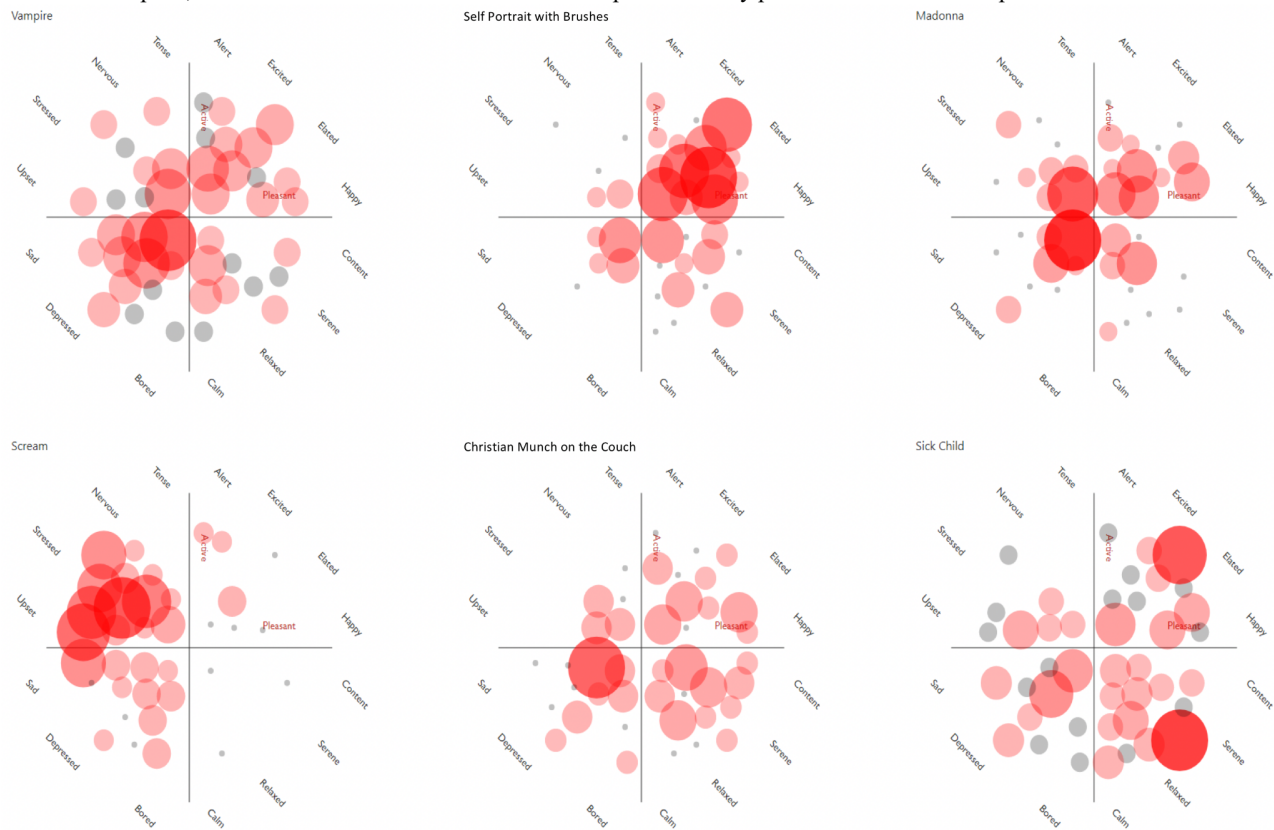

**Figure 5: Self-reported emotions for the six Munch paintings.**

In order to delve deeper into visitors' experiences, we now turn to our 39 semi-structured interviews. We note that 5 of our interviewees had a professional connection to the museum: 1 was a curator at the Munch Museum, and 4 others worked at collaborating institutions and had been invited by the museum to try out the experience. While the connection to the museum might potentially be a source of bias, the responses from these interviewees are interesting due to their professional expertise and insight into the subject matter of the exhibition as well as museum communication more broadly. We have noted this affiliation whenever these participants are quoted. The remaining 34 interviewees were self-selected public visitors, of which the majority were tourists visiting from abroad, representing a range of ages from 17 to 70, but with nearly half (16) being under 30 years old. 15 interviewees identified as female and 11 as male (the remainder were not asked about their gender).

## 5.1 Experiencing the paintings

The vast majority of the participants we interviewed (31 of 39) stated that they very much enjoyed the experience, describing it with words such as "very emotional", "touching", "surprising", "very meaningful", "very intense", "very impressive". During interviews, participants attributed their enjoyment to the enhanced emotional responses to the artwork provoked by the experience. For some, this enhancement came as a surprise: *"I was expecting more of a historical or intellectual sort of exercise but it became very emotional, and talking about feelings. I liked it a lot"* (P160).[1] And as P233 says, "*It was very touching for me. I never expected that.*"

In the interviews, participants shared stories of highly personal and emotional experiences. These seem to have been triggered through a combination of listening to the audio narrative, looking at the paintings and receiving questions to reflect upon: "*The picture of his father is a pretty distant picture when you just stand in front of it. It's not my father. It's not a person I know. It's just somebody reading a newspaper. But the way the questions were brought up made it very intense and very personal"* (P178, speaking about the painting Christian Munch on the Couch). For many, using headphones created a private space where this type of experience could take place. As P233 puts it, *"I could close myself in. That was very astonishing because normally with so many people around I can't concentrate very well and I feel kind of... um, going away. But in this particular time I was very calm and concentrated."*

Sensitive Pictures evidently encouraged participants to reflect on their emotions. The narrative prompts attached to each painting encouraged the fostering of personal connections between their lives and moments captured by Munch in his artworks, for example: "*You have to reflect not only on something outside yourself but you really have to go inside yourself as well and really connect with certain moments"* (P115, Museum Advisor). Such personal reflections allowed some users to better understand and process parts of their own life:

> Yeah actually I'm in a relationship right now and I'm thinking about it a lot because it's kind of difficult at the moment. (…) And yeah I mean this was about the difficulties of relationship and the two different sides that are in a relationship and it can be bad it can be good and you have to choose (...) all the emotions came back when I was staring at the picture on a painting. (P114 speaking about Vampire)

Participants generally demonstrated high levels of trust in the narrative content. Few questioned its validity, even of the content that was highly emotional. For example, at *The Scream*, the narrative draws connections between the piece and several socio-political topics such as climate change, war and disaster. Participants were generally willing to accept this interpretation of the piece, and use it to shape the ways they thought about and accessed the art:

> To me in that painting I kind of see... global warming and climate crisis and everything like that, and in the background of Scream, the world is like literally melting and you're falling into the sky. And humans are just like on a dark track into the abyss, and the figure is standing there and just in shock. (P211 talking about "Scream")

The vast majority of participants completed the self-reporting task for one or more of the art pieces, and many of the participants we interviewed enjoyed these questions as interesting prompts for reflection: "*I liked it. It asked questions that I wouldn't have considered otherwise. Interpretations of paintings that are different from my own*" (P211). However, this was not always a comfortable experience: "*I didn't expect the camera to turn on me and ask me these kinds of questions. (...) I'm surprised, a little bit shocked, a little bit hesitant but when I look back I'm not sure how comfortable I think it was"* (P115, Museum Advisor). Answering the questions frustrated some, who found it difficult to summarize their emotions in such a short space:

> I think it was very difficult because you could just type in one emotion or like one feeling. And I was standing there like, I don't know, roughly 2-3 minutes to think about one emotion that captures all the feelings I just felt staring at the picture. And I had a lot of feeling so I was like, I don't know what to type in. (P114)

---

[1] The participant numbering corresponds to the user id numbers generated by the backend system, which is why there are participant numbers that are higher than the total number of participants.

To others this challenge was actually enjoyable: "Normally I need more time to do that, to bring it to one word or two or three words. It's a little bit difficult, but it was also a good experience because I didn't expect me to react so fast. I felt proud of myself" (P233). Another noted that those questions that were more specific and well connected with the paintings worked best in triggering reflections on her own emotions.

It is notable that all but a few (5) of our interviewees expressed frustration with the self-reporting interface, especially with the affective sliders. This frustration was often attributed to the quantitative logic of the sliders, with participants saying that the options on the sliders did not necessarily capture the emotions provoked by the experience:

> It was hard sometimes to just have like binaries for the emotions, I think. To say if it either calms you or excites you. Especially because exciting... like they both seem really positive, I think, excitement and calmness. (P165)
>
> There was a bar, calm and exciting, wasn't it? (...) I don't know if that fits perfectly actually, because I was always like I'm excited about this feeling. Actually, I was never calm. (P112)

In contrast to the other visitors, the museum professionals we interviewed varied quite widely in their assessment of the experience. The participant who worked as a curator at the Munch Museum had a strong negative reaction, because she did not agree with the way in which the audio narratives presented the paintings. Another two museum professionals were also somewhat critical of the experience, indicating that they felt the audio narratives and questions were overly directive (a concern also shared by 3 other interviewees):

> P302 (Museum manager): But the thing I liked less was actually when they asked me questions directly and I had to respond to them. It felt that it is... not intrusive, but I think these paintings are so strong in their history. So I guess I have my own thoughts and I don't like to get directed.

However, the other museum professionals interviewed were more positive. A French museologist visiting on holiday (unaffiliated with the Munch Museum) explained enthusiastically that "*The experience is really interesting. Because it's different from the usual 'Munch was born in nanana and this painting is nanana'. It's more about feeling, emotion, and how you experience. It's really interesting*" (P65).

## 5.2 The "conversation" with Munch

The final part of the experience required users to talk to Munch on the phone. This part of the experience received a far more varied reception from participants, describing it with words such as "fun", "really cool", "intense", "very real", "creepy", "artificial", "Disney-esque", "weird", "a bit silly", "amazing", "touching", "corny" and "confusing". Some participants indicated that sitting down for a simulated "conversation" with the artist felt like a strong, emotional climax to their experience: "*Oh that was very touching. It was very touching. Yeah. It felt like you were talking to him*" (P81). Others indicated having more mixed feelings, sometimes wavering between fascination and not feeling able to take it fully seriously:

> I laughed a little because I'm not a really big believer in life after death you know. So he's coming back. You know it's a little corny to me at first but I accept it for a way of getting to where you, you know what you're trying to get. So I watch myself criticizing it and then but then go OK fine let's just let's do this. (P160)

Almost all of our interviewees (33 of 39) indicated that they indeed tried to answer Munch's questions by speaking out loud, even if some expressed discomfort with the fact that they were seated in a public space where other visitors could listen in as they passed by.

> Yeah I was a little nervous about that because, I don't know, sometimes it's a little bit scary, but it's very interesting especially because it was personalized. That was cool. I noticed that I was kind of always thinking about how I was showing emotions when I was doing it, but it was good. (P165)

However, a few indicated that they instead opted to answer inside their heads, or not at all; whereas some answered but in a less personal and emotional way than the other parts of the experience:

> I answered actually yes. Although I told Munch, thank you for the question but I think that that's not the moment to go into this conversation. But I will come back to it and I'm looking forward to our next phone call. That was my answer and it fitted perfectly into the time slot. (P178)

Several participants remarked that they felt odd or uncomfortable in answering, not due to the risk that others might hear their answers, but because they felt unease with the fact of speaking to a virtual character on a computer screen about their emotions. Some also expressed some confusion regarding the nature of the interaction, wondering about whether the fictional Munch character was meant to understand and react to their responses, and wondering how the system had identified the painting which they had indicated the strongest emotional response to.

## 5.3 The souvenir postcards

Most participants were very willing to reflect on their souvenir postcard and to explore the relationship between their subjective experience and the data it presented. This was true both of the self-reported data and the facial recognition data. Even while most participants expressed some confusion about how the data had been collected and what the visualization meant, the vast majority of participants agreed at least partly with the data presented: 11 participants stated clearly that they thought the data printed on the postcard represented what they had actually felt during the experience, whereas 19 partly agreed with the data, and only 2 participants stated that they were skeptical (for the remaining 7 we do not have data on this question).

> Yeah, I think it's true. According to this [indicating towards the self-reported data]: Very true. According to this, to the second [indicating towards the emotion detection data]: Hm. This is revealing I think, for me. Deep inside I think it's true. Yes. (P233)

Participants were often remarkably willing to offer interpretations to help make sense of the data presented to them. Some participants even changed their reflections upon examining their postcards, particularly in reference to the reading regarding frustration. Having previously stated that they were not frustrated, when seeing that the system indicated that they had experienced some level of frustration, several users attempted to find explanations for why this might have been true: *"I don't know about frustrated. Yeah I guess frustrated in the way that, you have to grapple with emotions as they come up and sometimes you don't want to"* (P211). One participant initially described the video conversation with Munch enthusiastically as *"very touching… It felt like you were talking to him… really lovely feeling"*, but when presented with the output from the emotion detection system saying she had been "very frustrated, very engaged and you enjoyed it somewhat", she reasoned: *"I think I was very engaged. I enjoyed it a lot. I wonder why it thought I was frustrated. My frustration was that it was a much smaller selection of works than I thought it would be"* (P81).

Some participants wavered, unable to decide whether to rely more on the emotion detection output or their own experience. In seeing explanations of the system's characterization compared to their own, a few users reasoned that even though they had enjoyed the experience, perhaps the system had picked up on underlying bodily sensations that were bothering them: "I was very engaged. Frustrated was probably more... I have back pain so maybe it caught that. I didn't feel frustrated. I think" (P165).

> [Interviewer:] This is what the camera saw in your face when you were watching the video. It says very frustrated, very engaged, and you enjoyed it somewhat.
> [Participant:] Yeah, okay. [Laughs] Very frustrated - that was because I had to pee! (P69)
> Some participants appeared to trust the facial recognition technology to accurately gauge their emotions more than they trusted their own self-reported data about their emotional state:
> Interviewer: Do you think this is actually what you felt? Is it true?
> Participant: Yes. Somehow frustrated. Yeah, engaged. Yeah I think it is. This [indicating towards the emotion detection data] looks more familiar to me than this [indicating towards the self-reported data].
> Interviewer: So the output from the camera, you believe that more than you believe the sliders?
> Participant: Yeah, yeah. (P115, Museum Advisor)

## 6 DISCUSSION

Following our approach of performance-led research in the wild, we now reflect on both the design and audience experience of Sensitive Pictures to draw out more general design knowledge. On balance, we believe that we succeeded in delivering an experience that engaged visitors with an overtly emotional interpretation of Munch's artworks and that was generally well received by both visitors (evidenced by their positive feedback) and the museum stakeholders (who decided to press ahead with a full-scale experience), though this was by no means universally the case, with several of each expressing important reservations. We reflect on three facets of emotional visiting experiences.

## 6.1 Provoking Emotional Interpretations

The provocation of emotional responses by layering new overtly emotional interpretations onto existing artworks appeared to be a successful element of the experience, setting an appropriately emotional tone for experiencing the paintings. Encouraging visitors to make connections to key personal relationships in their lives, while employing a deliberately intimate tone of voice, accompanying music, and use of headphones delivered an experience that some visitors described as being therapeutic, mirroring findings from previous experiences [63,72].

However, the approach was not universally appreciated. Some visitors found it to be too directive and had reservations about our approach to capturing and analyzing emotions that we consider below. Others found aspects of it to be uncomfortable, notably dealing with challenging emotional themes and talking aloud in the museum during the interview with Munch. Previous arguments for the deliberate use of discomfort in interaction design have stressed that introducing a degree of discomfort may help deliver enlightening experiences and that this might be social in nature (talking aloud in our case), based on control (being directive) or cultural (dealing with challenging themes), but also that such tactics needs to be carefully managed [7] as we also see here.

Curators wanted to balance emotional interpretation with explanatory information about the artist and their works. They were concerned about being overly directive. Some of these reservations were revealed and at least partially accommodated through the co-creation process. However, we need to recognise that emotional interpretations will need to be carefully introduced into many museums. In the near term, they might be offered as alternatives that sit alongside more conventional interpretations, but in the longer term it seems necessary to better justify them in the context of contemporary museum practice. One lens for viewing these kinds of emotional interpretations is that of "appropriation" - visitors are appropriating Munch's paintings for their own emotional purposes, while curators/artists are appropriating the paintings to provoke this. However, as discussed by Ryding at al. [63], while appropriation is often viewed in a positive light in HCI, as when users are able to tailor or adapt technologies to new purposes [15,19], it is a far more controversial idea in museums where, seen through a postcolonial lens, it raises the specter of misappropriating others' cultures [83]. Perhaps a more useful approach is to turn to ideas of re-enactment and re-interpretation that have emerged in museum practice and scholarship, driven by the challenges of documenting, preserving and exhibiting more ephemeral interactive works. Giannachi [29] notes that museums have long sought to "reactivate" exhibits whenever they bring them out of storage and exhibit them anew, and that this may involve strategies of re-enactment or re-mediation; she distinguishes the strategy of re-interpretation as interpreting existing works anew by drawing practices such as arrangement, homage, reappropriation, and emulation that are "commonly utilized in other disciplines". The idea of reactivation may provide a useful hook upon which to hang emotional interpretation.

## 6.2 Capturing emotions

While visitors often experienced strong emotions, capturing them proved to be more elusive. Both the self-reporting and computer vision techniques that we tried had limitations. Explicit self-rating of emotions proved challenging as visitors struggled to boil their emotions down to just a few words or to the simple rating scales of valence, arousal and control that underpin the Circumplex Model. The implicit capture of emotions through facial action detection also proved challenging given the inherent constraints of the software recognising a limited palette of emotions; the challenges of embedding cameras into the museum; and visitors' evident reluctance to express their inner feelings in this context. The hushed public setting of an art gallery is perhaps not a situation in which many people visibly emote, and even though we tried to get them to talk as part of an interview, the emotions that they appear to have felt internally did not always break the surface to the point of being visibly detectable.

There might be various strategies for trying to better capture emotion in future experiences. Regarding self-reporting, one might attempt to improve the self-reporting mechanisms by using alternative or multiple models of emotion (e.g. drawing on Ekman's model of basic emotions [23]). Alternatively, one might attempt to employ more powerful tools to elicit stories such as the tangible objects developed as part of Isbister et al.'s Sensual Evaluation Instrument [39].

Designers might attempt to create experiences in which visitors are further encouraged to perform or even exaggerate their emotions, perhaps through even more dramatized exchanges or by placing the interaction in a dedicated room away from other visitors (e.g. [73]), and of course we might expect computer vision to become more sensitive given further development.

Based on our experience, we commend the strategy of deploying multiple techniques within a single experience so that each is able to contribute some data. Interestingly, ours is not an approach of data fusion in which we are trying to integrate multiple sources of data to reach a common consensus, but rather perhaps one of "data diffusion" in which each technique yields a different, perhaps small, insight into emotions that can then be juxtaposed with others during later reflection. We suggest that this might be achieved by deploying the different techniques in sequence (as in our design) rather than in parallel, tipping the visitor back and forth between experiencing a provocation and describing their emotional responses in different ways, before ultimately trying to reconcile the different data at the end.

## 6.3    Reflecting on emotions

While somewhat difficult to capture, the data that visitors were able to provide was powerful for provoking reflections on emotional experience when presented back to them. It appears that showing visitors even simple representations of their own data can be a powerful stimulus to storytelling, with people keen to make their own interpretations.

In line with previous research into data souvenirs, the postcards appeared to be an effective way of achieving this. A notable feature of their design was the degree of ambiguity with which they presented visitor's emotion data. On the one hand they protracted the outputs of the computer vision system in a fuzzy and ambiguous way that invited interpretation, conveyed through phrasing such as "seems like", "somewhat" and the question "Do you think this is actually what you felt?" shown in Figure 4. On the other hand, they included a precise presentation of self-reported emotions in terms of the discrete points plotted on the Circumplex Model along with the subsequent statements about how they felt at each painting. The data on the postcards appears to invite people to make an interpretation, affirming previous arguments by Sengers et al. [65] that ambiguity can be a means of provoking interpretation, and the specific tactics espoused by Gaver et al. [28] in which ambiguity can be conveyed by presenting information in ways that are both fuzzy and overly precise.

What most stands out from our findings is how visitors tended to construct post hoc rationalizations of their emotional experience that agreed with, or at least accommodated, the "results" reported by the system, even when this differed from their initial reflections. In spite of revealing some skepticism towards the general idea that systems can understand human emotions when asked in the abstract, in practice there was a striking tendency to reconcile their accounts with the system – bluntly, to agree with it. We can speculate as to various reasons for this.

Of course, visitors did not experience, describe, or rationalize their emotions in a vacuum. Museums are inherently social institutions, with museum staff and other visitors often nearby, especially in a popular location such as the Munch Museum. Interviews were also conducted in person, creating a social situation with its own norms: for example, interviewees rarely express negative opinions of an interactive experience such as Sensitive Pictures in extreme terms in much the same way as they would rarely knowingly offend a stranger by insulting their appearance.

Our design also included mechanisms that tried to win visitors' trust, most notably the "data token." We were surprised that nearly all the participants were comfortable donating their data, suggesting a high level of trust in the experience in general. It may also be that visitors were trusting in the apparently scientific nature of the technology and presentation of their data (the plots mentioned previously), mirroring claims from studies in science communication that have suggested that presentation formats that appear scientific – e.g. using scientific language and charts – may increase participants' belief in a message [20,31,54,76]. Finally, we note that invoking "ambiguity of relationship" and "ambiguity of context" are further tactics (in addition to the "ambiguity of information" noted earlier) for creating ambiguity as proposed by Gaver et al [28]; perhaps our visitors were creating interpretations that could resolve their ambiguous relationship to the system and its ambiguous presence within art gallery. Unpacking the reasons for visitors constructing accounts that tend to agree with the system is clearly a topic for further research.

We draw attention to further opportunities for displaying emotion data in museums beyond take-away souvenirs. Boehner et al. [13], for example, describe the deployment of "ambient displays of Affective Interaction" positioned around the museum, which might present more complex visualizations that could not fit onto the back of a postcard. Personal mobile devices or stations around the museum might allow for interactive visualizations. We might even consider other opportunities for "paper displays" such as extending exhibit labels, posters or catalogs. For example, the kinds of visualizations we introduced in Figure 5 might be displayed next to paintings as "emotional labels", complementing more traditional ones, or be delivered as interactive augmented reality overlays so the visitor could highlight their own response in relation to others.

In summary, ambiguous presentations of personal data through tangible data souvenirs appear to be powerful stimuli for eliciting emotional storytelling among visitors, but also come with considerable risk given a possible tendency to agree with the system's view. We recommend harnessing this approach with some subtlety, for example carefully framing visitor experiences to distinguish between storytelling and claims to "truth", or finding more nuanced ways of displaying emotion data within the museum.

## 6.4 Emotional trajectories

A notable feature of Sensitive Pictures is how it combined various emotional provocations and reflections into an overall visitor experience, with the visitor moving back and forth between provocation and reflection multiple times. We conclude our discussion by reflecting on how the component parts were integrated to create a more holistic emotional experience.

One lens for viewing holistic user journeys through interactive experiences, especially cultural ones, is that of trajectories. Such experiences can be viewed as interleavings of canonical trajectories that express the experience as designed, participant trajectories that express it as encountered, and historic trajectories that consider it as subsequently recounted [6]. The notion of "affect trajectories" as a potential concept for unfolding and highlighting visitor's reactions to unsettling or emotionally challenging interactive museum experiences has previously been discussed in [62]. Other relevant work has highlighted how "soma trajectories" through embodied experiences may be designed to oscillate between comfortable and uncomfortable, familiar and strange, and inside and outside perspectives [77], mirroring a wider recognition of the importance of both participants and designers flipping between first, second and third person perspectives [37]. These notions of oscillating trajectories and flipping perspectives are mirrored in the design of Sensitive Pictures. Our design encourages participants to move back and forth between different emotional provocations (experiencing up to six paintings and the conversation with Munch) that are interspersed with different modes of emotional reflection (self reporting via scales and questions, and Munch's questions). Individual participant trajectories may vary within this general design, for example in the number and order of paintings encountered, while the idea of the historic trajectories is supported through the postcard that encourages reflection and recounting.

We draw attention to elements of the design that serve to integrate the various components into a whole: the data that is captured throughout is presented on the postcard at the end; the strongest self-reported response to the paintings shapes the interview with Munch; and of course, there is the enduring presence of Munch and his paintings in the coherent setting of the museum that integrates the experience at a narrative level.

Our design might be described as following an *emotional trajectory*, one that oscillates between different emotional provocations, reflections and data capture techniques, and that draws them together at the end in a final souvenir. Such an approach allows participants to repeatedly sample their own emotions and can accommodate multiple affective technologies and potentially different perspectives on and/or models of emotion (although we used the Circumplex Model throughout, we could have introduced other models such as Ekman's proposed universal or basic emotions).

Trajectories as a framework has long been used to describe physical motion through time and space, whether geographically or through sequences of selection and attention through game play or narrative. Emotions are implicated in the dynamic, artistic interactions that the trajectories framework is often applied to, yet in Sensitive Pictures we see a clear case for emotions to be held up for analysis in their own right. The project involved multiple types of touchpoint and asked visitors to engage in varying ways that were ultimately outside the direct control of the designers. The canonical trajectory – i.e. the one the designers envisioned a "typical" (or perhaps ideal) visitor having – shows a regular alternation between the visitor experiencing Munch (his works followed by his image and "voice") and the visitor experiencing their own feelings about Munch's work. This emotional back-and-forth between Munch's work and their own emotions is paralleled by their attention being directed first outwards, to experience the paintings around them, and then inwards for self-reflection as prompted and enabled by the visitor's smartphone, held close to their body.

We can also reflect on our various data to consider the general shape of the participant trajectory, i.e, how participants actually encountered Sensitive Pictures in practice. Figure 6 shows our post-hoc attempt to approximate this participant trajectory as a reflection on our design. The shapes of the curves here are intended to broadly convey how we think participants encountered the different elements of the experience. Their relative heights reflect estimates of the levels of emotional provocation and reflection that participants experienced: our interviewees tended to describe strong emotional reactions to the paintings, but were more mixed in their responses to the conversation with Munch; while their self-reported emotional responses (see beginning of Section 5) indicate that the different paintings tended to provoke different levels of emotional response. The durations of the various encounters along the horizontal axis reflect the time spent on the different parts of the experience. While individual participants chose to encounter the paintings in different orders, we have ordered the curves based on the overall popularity of each as the first experience chosen. We have also labeled how the different kinds of data that were gathered (self-report data and facial action analysis data) served to connect together the overall journey.

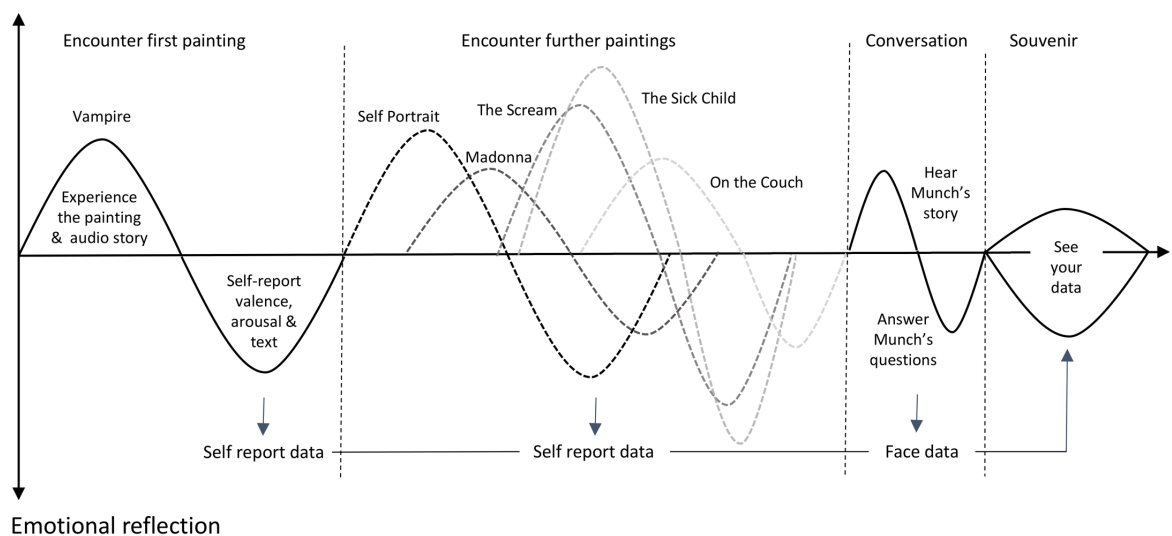

**Figure 6:** Approximation of the Participant Emotional Trajectory through Sensitive Pictures. The height of the curve for each painting reflects the number of participants that gave the strongest response to that painting in their self-report data.

## 7 CONCLUSIONS

Overall, we propose that we were successful in designing an "emotional visitor experience" for the Munch Museum, by which we mean one that provoked an emotional response to Munch's paintings and encouraged visitors to reflect on this. The experience was well received by the public, although it gained a more mixed response from curators, suggesting that more work needs to be done to situate the approach within contemporary museum practice and/or that such experiences might initially find a home in museums that offer overtly "alternative" interpretations alongside more conventional ones.

We draw attention to the strategy of interleaving overt emotional provocations with both self-reported and machine-reported capture of emotional data and tangible souvenirs that encourage reflection on both. Provocations that were based on carefully acted stories delivered as an intimate audio experience appeared to work well. The capture of emotion data was more challenging, however, as visitors struggled to boil complex emotions down to simple scales and words and did not always visibly emote their inner feelings to the computer vision software. The ambiguous portrayal of the data on the souvenir postcards appeared to demand interpretation. However, we were surprised that participants often seemed willing to accept the system's presentation as a truthful measure of their emotions, sometimes offering interpretations to help make the data "fit", suggesting a need to treat the approach with caution and better reveal the challenges and nuances of emotion data.

We conclude by revisiting the wider debate in HCI concerning the complex and overlapping relationships between Affective Computing, Affective Interaction and Technology as Experience. Overall, we designed an emotional user experience that combined aspects of Affective Interaction and Affective Computing. Affective Interaction's approach of humans constructing interpretations of emotions, enabled by digital technologies, is clearly a core part of our experience. However, Affective Computing's core idea that computers might interpret emotions is also present in our experience in the attempt to apply models of emotions to facial movements detected by video cameras.

On reflection, it seems to us that the key difference between Affective Interaction and Affective Computing lies in what might be called the "locus of interpretation", that is in whether it is humans or computers that are interpreting emotions. Sensitive Pictures employs both, with humans and computers interpreting emotions at different points along an overall experience that, more widely, aims to generate as well as understand emotional experiences. In this case, Artificial Intelligence provides a mirror to human emotions; it is a way of reflecting back to us something we do not normally see for ourselves. Moreover, AI can be seen as a distorted mirror, in the sense that it sees our emotions somehow differently. We suggest that this opens up a potentially powerful role for Affective Computing: reflecting unusual views of emotions back to humans to help them better understand their own emotional responses and so undertake more emotionally intelligent actions. However, the propensity for people to trust in technology and perhaps also institutions like museums means that this needs to be treated carefully. For those who are trying to develop Affective Computing to take decisions for humans or directly adapt to them, we suggest that the kinds of emotional experience we have outlined here may help reveal how humans understand and talk about their own emotional responses, which might perhaps shape how to present the actions of Affective Computing back to them.

## ACKNOWLEDGEMENTS

We gratefully acknowledge the support of the European Union for the Meaningful Personalization of Hybrid Virtual Museum Experiences Through Gifting and Appropriation (GIFT) project (Grant ID: 727040), and of the Engineering and Physical Sciences research Council (EPSRC) for the EPSRC Centre for Doctoral Training in Horizon: Creating Our Lives in Data (Grant ID: EP/S023305/1) and Horizon: Trusted Data-Driven Products (Grant ID: EP/T022493/1) projects.## REFERENCES

[1] Ryan Aipperspach, Ben Hooker, and Allison Woodruff. 2011. Data Souvenirs: Environmental psychology and reflective design. International Journal of Human-Computer Studies 69, 5: 338–349. https://doi.org/10.1016/j.ijhcs.2010.12.003

[2] Nalini Ambady and Robert Rosenthal. 1992. Thin slices of expressive behavior as predictors of interpersonal consequences: A meta-analysis. Psychological Bulletin 111, 2: 256–274. https://doi.org/10.1037/0033-2909.111.2.256

[3] Silke Arnold-de Simine. 2013. Mediating Memory in the Museum: Trauma, Empathy, Nostalgia. Palgrave Macmillan, London, UK. Retrieved from https://doi.org/10.1057/9781137352644

[4] Gaynor Bagnall. 2003. Performance and performativity at heritage sites. Museum and Society 1, 2: 87–103. https://doi.org/10.29311/mas.v1i2.17

[5] Genevieve Bell. 2002. Making Sense of Museums: The Museum as 'Cultural Ecology.' Retrieved June 3, 2021 from http://echo.iat.sfu.ca/library/bell_02_museum_ecology.pdf

[6] Steve Benford and Gabriella Giannachi. 2011. Performing Mixed Reality. The MIT Press, Cambridge, MA, London, U.K.

[7] Steve Benford, Chris Greenhalgh, Gabriella Giannachi, Brendan Walker, Joe Marshall, and Tom Rodden. 2012. Uncomfortable interactions. In Proceedings of the SIGCHI Conference on Human Factors in Computing Systems (CHI '12), 2005–2014. https://doi.org/10.1145/2207676.2208347

[8] Todd Bentley, Lorraine Johnston, and Karola von Baggo. 2005. Evaluation using cued-recall debrief to elicit information about a user's affective experiences. In Citizens Online: Considerations for Today and the Future, 1–10.

[9] Alberto Betella and Paul F. M. J. Verschure. 2016. The Affective Slider: A Digital Self-Assessment Scale for the Measurement of Human Emotions. PLOS ONE 11, 2: e0148037. https://doi.org/10.1371/journal.pone.0148037

[10] Graham Black (ed.). 2020. Museums and the Challenge of Change: Old Institutions in a New World. Routledge.

[11] Kirsten Boehner, Rogério DePaula, Paul Dourish, and Phoebe Sengers. 2005. Affect: From Information to Interaction. In Between sense and sensibility.

[12] Kirsten Boehner, Rogério DePaula, Paul Dourish, and Phoebe Sengers. 2007. How emotion is made and measured. International Journal of Human-Computer Studies 65, 275–291.

[13] Kirsten Boehner, Phoebe Sengers, and Geri Gay. 2006. Affective presence in museums: Ambient systems for creative expression. Digital Creativity 16, 2: 79–89.20

[14] Margaret M. Bradley and Peter J. Lang. 1994. Measuring emotion: The self-assessment manikin and the semantic differential. Journal of Behavior Therapy and Experimental Psychiatry 25, 1: 49–59. https://doi.org/10.1016/0005-7916(94)90063-9

[15] Jennie Carroll, Steve Howard, Frank Vetere, Jane Peck, and John Murphy. 2001. Identity, Power And Fragmentation in Cyberspace: Technology Appropriation by Young People. ACIS 2001 Proceedings. Retrieved from https://aisel.aisnet.org/acis2001/6

[16] Antonio R. Damasio. 1995. Descartes' Error: Emotion, Reason, and the Human Brain. Harper Perennial.

[17] M. Ali Akber Dewan, Mahbub Murshed, and Fuhua Lin. 2019. Engagement detection in online learning: a review. Smart Learning Environments 6, 1: 1. https://doi.org/10.1186/s40561-018-0080-z

[18] John Dewey. 2005. Art as Experience. Penguin.

[19] Paul Dourish. 2001. Where the Action Is: The Foundations of Embodied Interaction. The MIT Press, Cambridge, MA.

[20] Pierre Dragicevic and Yvonne Jansen. 2018. Blinded with Science or Informed by Charts? A Replication Study. IEEE Transactions on Visualization and Computer Graphics 24, 1: 781–790. https://doi.org/10.1109/TVCG.2017.2744298

[21] J. O. Egede, S. Song, T. A. Olugbade, C. Wang, ACDC Williams, H. Meng, M. Aung, N. D. Lane, M. Valstar, and N. Bianchi-Berthouze. 2020. EMOPAIN Challenge 2020: Multimodal Pain Evaluation from Facial and Bodily Expressions. In 2020 15TH IEEE INTERNATIONAL CONFERENCE ON AUTOMATIC FACE AND GESTURE RECOGNITION (FG 2020), 849–856. Retrieved September 9, 2021 from https://discovery.ucl.ac.uk/id/eprint/10116032/

[22] Lina Eklund. 2020. A Shoe Is a Shoe Is a Shoe: Interpersonalization and Meaning-making in Museums – Research Findings and Design Implications. International Journal of Human-Computer Interaction 36, 16: 1503–1513. https://doi.org/10.1080/10447318.2020.1767982

[23] Paul Ekman. 1999. Basic emotions. In Handbook of cognition and emotion. John Wiley & Sons Ltd, New York, NY, US, 45–60. https://doi.org/10.1002/0470013494.ch3

[24] Paul Ekman and Wallace V. Friesen. 1978. Facial Action Coding System: A Technique for the Measurement of Facial Movement. Consulting Psychologists Press', Washington, DC.

[25] John H. Falk and Lynn Diane Dierking. 2012. The Museum Experience Revisited. Routledge, London, England.

[26] Lesley Fosh, Steve Benford, Stuart Reeves, Boriana Koleva, and Patrick Brundell. 2013. 'See Me, Feel Me, Touch Me, Hear Me': Trajectories and Interpretation in a Sculpture Garden. In CHI 2013: Changing Perspectives, 149–158.

[27] William Gaver. 2009. Designing for emotion (among other things). Philosophical Transactions of The Royal Society B Biological Sciences 364, 3597–3604.

[28] William Gaver, Jacob Beaver, and Steve Benford. 2003. Ambiguity as a Resource for Design. 233–240. Retrieved July 31, 2017 from https://dl.acm.org/doi/10.1145/642611.642653

[29] Gabriella Giannachi. 2017. At the edge of the 'living present': re-enactments and re-interpretations as strategies for the preservation of performance and new media art. In Histories of Performance Documentation: Museum, Artistic, and Scholarly Practices, Gabriella Giannachi and Jonah Westerman (eds.). Routledge, 115–131.

[30] Joseph F. Grafsgaard, Joseph B. Wiggins, Alexandria Katarina Vail, Kristy Elizabeth Boyer, Eric N. Wiebe, and James C. Lester. 2014. The Additive Value of Multimodal Features for Predicting Engagement, Frustration, and Learning during Tutoring. In Proceedings of the 16th International Conference on Multimodal Interaction (ICMI '14), 42–49. https://doi.org/10.1145/2663204.2663264

[31] Jenifer Haard, Michael D. Slater, and Marilee Long. 2004. Scientese and ambiguous citations in the selling of unproven medical treatments. Health Communication 16, 4: 411–426. https://doi.org/10.1207/s15327027hc1604_2

[32] Marek Hatala and Ron Wakkary. 2005. Ontology-based user modeling in an augmented audio reality system for museums. UMUAI 15, 3–4: 339–380.

[33] Adrian Hazzard, Steve Benford, and Gary Burnett. 2015. Sculpting a mobile music soundtrack. In SIGCHI Conference on Human Factors in Computing Systems, 387–396.

[34] Kristina Höök. 2006. Designing familiar open surfaces. In Fourth Nordic Conference on Human-Computer Interaction, 242–251.

[35] Kristina Höök. 2009. Affective loop experiences: designing for interactional embodiment. Philosophical Transactions of the Royal Society B: Biological Sciences 364, 1535: 3585–3595. https://doi.org/10.1098/rstb.2009.0202

[36] Kristina Höök. 2012. Affective Computing. In The Encyclopedia of Human-Computer Interaction (2nd ed.). Interaction Design Foundation.

[37] Kristina Höök, Steve Benford, Paul Tennent, Vasiliki Tsaknaki, Miquel Alfaras, Juan Martinez Avila, Christine Li, Joseph Marshall, Claudia Daudén Roquet, Pedro Sanches, Anna Ståhl, Muhammad Umair, Charles Windlin, and Feng Zhou. 2021. Unpacking Non-Dualistic Design: The Soma Design Case. ACM Transactions on Computer-Human Interaction 28, 6: 40:1-40:36. https://doi.org/10.1145/3462448

[38] Kristina Höök, Anna Ståhl, Petra Sundström, and Jarmo Laaksolahti. 2008. Interactional empowerment. In ACM CHI 2008 Conference on Human Factors in Computing Systems, 647–656.

[39] Katherine Isbister, Kristina Höök, Michael Sharp, and Jarmo Laaksolahti. 2006. The Sensual Evaluation Instrument: Developing an Affective Evaluation Tool. In Proceedings of the SIGCHI Conference on Human Factors in Computing Systems (CHI '06), 1163–1172. https://doi.org/10.1145/1124772.1124946

[40] Jack Katz. 1999. How Emotions Work. University of Chicago Press.

[41] Jenny Kidd. 2018. "Immersive" heritage encounters. The Museum Review 3, 1.

[42] Lucian Leahu, Steve Schwenk, and Phoebe Sengers. 2008. Subjective objectivity: negotiating emotional meaning. In Proceedings of the 7th ACM conference on Designing interactive systems (DIS '08), 425–434. https://doi.org/10.1145/1394445.1394491

[43] Joseph Ledoux. 1996. The Emotional Brain: The mysterious underpinnings of emotional life. Simon and Schuster.

[44] Ruth Leys. 2011. The Turn to Affect: A Critique. Critical Inquiry 37, 3: 434–472. https://doi.org/10.1086/659353

[45] Madelene Lindström, Anna Ståhl, Kristina Höök, Petra Sundström, Jarmo Laaksolathi, Marco Combetto, Alex Taylor, and Roberto Bresin. 2006. Affective diary: designing for bodily expressiveness and self-reflection. In CHI '06 Extended Abstracts on Human

Factors in Computing Systems. Association for Computing Machinery, New York, NY, USA, 1037–1042. Retrieved September 2, 2021 from https://doi.org/10.1145/1125451.1125649

[46] Brais Martinez, Michel F. Valstar, Bihan Jiang, and Maja Pantic. 2019. Automatic Analysis of Facial Actions: A Survey. IEEE Transactions on Affective Computing 10, 3: 325–347. https://doi.org/10.1109/TAFFC.2017.2731763

[47] John McCarthy and Peter Wright. 2004. Technology as Experience. The MIT Press, Cambridge, MA. Retrieved August 9, 2017 from https://www.researchgate.net/publication/224927635_Technology_as_Experience

[48] Bethany McDaniel, S. D'Mello, Brandon G. King, P. Chipman, Kristy M. Tapp, and A. Graesser. 2007. Facial Features for Affective State Detection in Learning Environments. In Proceedings of the Annual Meeting of the Cognitive Science Society. Retrieved September 6, 2021 from https://www.semanticscholar.org/paper/Facial-Features-for-Affective-State-Detection-in-McDaniel-D%27Mello/6acd50c5c98d68f6294e89032291f1463e7fb26e

[49] Anthony D. McDonald, Farzan Sasangohar, Ashish Jatav, and Arjun H. Rao. 2019. Continuous monitoring and detection of post-traumatic stress disorder (PTSD) triggers among veterans: A supervised machine learning approach. IISE Transactions on Healthcare Systems Engineering 9, 3: 201–211. https://doi.org/10.1080/24725579.2019.1583703

[50] Andrew McStay. 2018. Emotional AI: The Rise of Empathic Media. SAGE. Retrieved November 18, 2020 from https://uk.sagepub.com/en-gb/eur/emotional-ai/book251642

[51] Oonagh Murphy. 2019. The changing shape of museums in an increasingly digital world. In Connecting Museums, Mark O'Neill and Glenn Hooper (eds.). Routledge, London, UK.

[52] Christian Nold (ed.). 2009. Emotional Cartography - Technologies of the Self. emotionalcartography.net/. Retrieved from http://emotionalcartography.net/

[53] Elena Not, Massimo Zancanaro, Mark T. Marshall, Daniela Petrelli, and Anna Pisetti. 2017. Writing Postcards from the Museum: Composing Personalised Tangible Souvenirs. In Proceedings of the 12th Biannual Conference on Italian SIGCHI Chapter (CHItaly '17), 1–9. https://doi.org/10.1145/3125571.3125583

[54] Anshul Vikram Pandey, Anjali Manivannan, Oded Nov, Margaret Satterthwaite, and Enrico Bertini. 2014. The Persuasive Power of Data Visualization. IEEE Transactions on Visualization and Computer Graphics 20, 12: 2211–2220. https://doi.org/10.1109/TVCG.2014.2346419

[55] Daniela Petrelli, Mark T. Marshall, Sinéad O'Brien, Patrick McEntaggart, and Ian Gwilt. 2017. Tangible Data Souvenirs as a Bridge between a Physical Museum Visit and Online Digital Experience. Personal and Ubiquitous Computing 21, 2: 281–295. https://doi.org/10.1007/s00779-016-0993-x

[56] Daniela Petrelli and Elena Not. 2005. User-centred design of flexible hypermedia for a mobile guide: reflections on the hyperaudio experience. User Modeling and User-Adapted Interaction 15, 3–4: 303–338.

[57] R Picard. 1997. Affective Computing. MIT Press, Cambridge.

[58] Fabien Ringeval, Björn Schuller, Michel Valstar, Nicholas Cummins, Roddy Cowie, Leili Tavabi, Maximilian Schmitt, Sina Alisamir, Shahin Amiriparian, Eva-Maria Messner, Siyang Song, Shuo Liu, Ziping Zhao, Adria Mallol-Ragolta, Zhao Ren, Mohammad Soleymani, and Maja Pantic. 2019. AVEC 2019 Workshop and Challenge: State-of-Mind, Detecting Depression with AI, and Cross-Cultural Affect Recognition. In Proceedings of the 9th International on Audio/Visual Emotion Challenge and Workshop (AVEC '19), 3–12. https://doi.org/10.1145/3347320.3357688

[59] Ognjen Rudovic, Jaeryoung Lee, Lea Mascarell-Maricic, Björn W. Schuller, and Rosalind W. Picard. 2017. Measuring Engagement in Robot-Assisted Autism Therapy: A Cross-Cultural Study. Frontiers in Robotics and AI 4: 36. https://doi.org/10.3389/frobt.2017.00036

[60] James A. Russell. 1980. A circumplex model of affect. Journal of Personality and Social Psychology 39, 6: 1161–1178. https://doi.org/10.1037/h0077714

[61] Karin Ryding and Jonas Fritsch. 2020. Play Design as a Relational Strategy to Intensify Affective Encounters in the Art Museum. In Proceedings of the 2020 ACM on Designing Interactive Systems Conference (DIS '20), 681–693. https://doi.org/10.1145/3357236.3395431

[62] Karin Ryding and Anders Sundnes Løvlie. 2018. Monuments For A Departed Future: Designing For Critical Engagement With An Ideologically Contested Museum Collection. In MW18: Museums and the Web 2018. Retrieved from https://mw18.mwconf.org/paper/monuments-for-a-departed-future-designing-for-critical-engagement-with-an-ideologically-contested-museum-collection/

[63] Karin Ryding, Jocelyn Spence, Anders Sundnes Løvlie, and Steve Benford. 2021. Interpersonalizing Intimate Museum Experiences. International Journal of Human–Computer Interaction: 1–22. https://doi.org/10.1080/10447318.2020.1870829

[64] Holger Schnädelbach, Stefan Rennick Egglestone, Stuart Reeves, Steve Benford, Brendan Walker, and Michael Wright. 2008. Performing thrill: designing telemetry systems and spectator interfaces for amusement rides. In Proceedings of the SIGCHI Conference on Human Factors in Computing Systems (CHI '08), 1167–1176. https://doi.org/10.1145/1357054.1357238

[65] Phoebe Sengers and Bill Gaver. 2006. Staying open to interpretation: engaging multiple meanings in design and evaluation. In Proceedings of the 6th conference on Designing Interactive systems (DIS '06), 99–108. https://doi.org/10.1145/1142405.1142422

[66] Nina Simon. 2010. The Participatory Museum. Museum 2.0, Santa Cruz, California. Retrieved January 27, 2016 from http://www.participatorymuseum.org/

[67] Laurajane Smith. 2014. Changing Views? Emotional Intelligence, Registers of Engagement and the Museum Visit. In Museums as Sites of Historical Consciousness: Perspectives on museum theory and practice in Canada, V. Gosselin and P. Livingstone (eds.). UBC Press, Vancouver.

[68] Laurajane Smith. 2015. Theorizing Museum and Heritage Visiting. In The International Handbooks of Museum Studies, Volume 1: Museum Theory, Kylie Message and Andrea Witcomb (eds.). John Wiley & Sons, 459–484. https://doi.org/10.1002/9781118829059.wbihms122

[69] Laurajane Smith. 2021. Emotional Heritage: Visitor Engagement at Museums and Heritage Sites. Routledge. Retrieved September 8, 2021 from https://www.routledge.com/Emotional-Heritage-Visitor-Engagement-at-Museums-and-Heritage-Sites/Smith/p/book/9781138888654

[70] Laurajane Smith and Gary Campbell. 2015. The Elephant in the Room: Heritage, Affect, and Emotion. In A Companion to Heritage Studies, William Logan, Máiréad Nic Craith and Ullrich Kockel (eds.). Wiley-Blackwell, 443–460. Retrieved from https://onlinelibrary.wiley.com/doi/abs/10.1002/9781118486634.ch30

[71] Laurajane Smith, Margret Wetherell, and Gary Campbell (eds.). 2018. Emotion, Affective Practices, and the Past in the Present. Routledge.

[72] Jocelyn Spence, Benjamin Bedwell, Michelle Coleman, Steve Benford, Boriana N. Koleva, Matt Adams, Ju Row Farr, Nick Tandavanitj, and Anders Sundnes Løvlie. 2019. Seeing with New Eyes: Designing for In-the-Wild Museum Gifting. In Proceedings of ACM CHI Conference on Human Factors in Computing Systems (CHI 2019). https://doi.org/10.1145/3290605.3300235

[73] Jocelyn Spence, Dimitrios Paris Darzentas, Yitong Huang, Harriet R. Cameron, Eleanor Beestin, and Steve Benford. 2020. VRtefacts: Performative Substitutional Reality with Museum Objects. In Proceedings of the 2020 ACM Designing Interactive Systems Conference (DIS '20), 627–640. https://doi.org/10.1145/3357236.3395459

[74] Oliviero Stock, Massimo Zancanaro, Paolo Busetta, Charles Callaway, Antonio Krüger, Michael Kruppa, Tsvi Kuflik, Elena Not, and Cesare Rocchi. 2007. Adaptive, intelligent presentation of information for the museum visitor in PEACH. User Modeling and User-Adapted Interaction 17, 3: 257–304. https://doi.org/10.1007/s11257-007-9029-6

[75] Petra Sundström, Anna Ståhl, and Kristina Höök. 2007. In situ informants exploring an emotional mobile messaging system in their everyday practice. International Journal of Human-Computer Studies 65, 388–403.

[76] Aner Tal and Brian Wansink. 2016. Blinded with science: Trivial graphs and formulas increase ad persuasiveness and belief in product efficacy. Public Understanding of Science 25, 1: 117–125. https://doi.org/10.1177/0963662514549688

[77] Paul Tennent, Kristina Höök, Steve Benford, Vasiliki Tsaknaki, Anna Ståhl, Claudia Dauden Roquet, Charles Windlin, Pedro Sanches, Joe Marshall, Christine Li, Juan Pablo Martinez Avila, Miquel Alfaras, Muhammad Umair, and Feng Zhou. 2021. Articulating Soma Experiences using Trajectories. In Proceedings of the 2021 CHI Conference on Human Factors in Computing Systems, 1–16. Retrieved December 10, 2021 from https://doi.org/10.1145/3411764.3445482

[78] Michel Valstar. 2014. Automatic Behaviour Understanding in Medicine. In Proceedings of the 2014 Workshop on Roadmapping the Future of Multimodal Interaction Research including Business Opportunities and Challenges (RFMIR '14), 57–60. https://doi.org/10.1145/2666253.2666260

[79] Vassilios Vlahakis, Thomas Pliakas, Athanasios M. Demiris, and Nikolaos Ioannidis. 2003. Design and Application of an Augmented Reality System for continuous, context-sensitive guided tours of indoor and outdoor cultural sites and museums. In VAST, 155–164. Retrieved January 30, 2017 from http://lifeplus.miralab.unige.ch/html/papers/LIFEPLUS-VAST2003-revised.pdf

[80] Margret Wetherell. 2012. Affect and Emotion: a New Social Science Understanding. Sage, Los Angeles, London.

[81] Margret Wetherell. 2015. Trends in the Turn to Affect: A Social Psychological Critique. Body & Society 21, 2: 139–166. https://doi.org/10.1177/1357034X14539020

[82] Andrea Witcomb. 2013. Understanding the role of affect in producing a critical pedagogy for history museums. Museum Management and Curatorship 28, 3: 255–271. https://doi.org/10.1080/09647775.2013.807998

[83] Bruce H. Ziff and Pratima V. Rao (eds.). 1997. Borrowed Power: Essays on Cultural Appropriation. Rutgers University Press.